\documentclass{aa} % for a paper on 2 column

\usepackage{graphicx}

\usepackage{mathptmx}    
\usepackage{latexsym,amsmath,amssymb,natbib}
\usepackage{graphicx}% Include figure files
\usepackage{dcolumn}% Align table columns on decimal point
\usepackage{bm}% bold math
\usepackage{hyperref}% add hypertext capabilities
\usepackage{ulem}
\usepackage{txfonts}
\usepackage{textcomp}
\usepackage{amsmath}

\title{Observation of bottom-up formation for charged grain aggregates related to pre-planetary evolution
beyond the bouncing barrier
}

\begin{document}

\titlerunning{Formation of charged grain aggregates}

\author{ Felix Jungmann     \inst{1}       \and
        Gerhard Wurm    \inst{1}\\ 
}

\institute{   University of Duisburg-Essen\\ 
              Faculty of Physics \\
              Lotharstr. 1-21\\
              47057 Duisburg \\
              \email{felix.jungmann@uni-due.de}           %  \\
%             \emph{Present address:} of F. Author  %  if needed
}

   \date{--;--}

% \abstract{}{}{}{}{} 
% 5 {} token are mandatory
 
  \abstract
  % context heading (optional)
  % {} leave it empty if necessary
   {Particles in protoplanetary disks go through a number of phases that are dominated by collisions. In each of these events, grains exchange electrical charge via triboelectric effects. This enhances the stability of particle aggregates.}
  % aims heading (mandatory)
   {Dielectric grains are easily charged by collisions. Here, we investigate whether a charge is capable of inducing an aggregation of particles and we consider how collision properties, such as sticking velocities and collisional cross-sections, are altered. }
  % methods heading (mandatory)
   {We explored aggregation in microgravity experiments based on the observation of the motion of submillimeter (submm) grains following many collisions. In the process, grains attract each other, collide, stick, and ultimately form small aggregates.}
  % results heading (mandatory)
   {We observed a bottom-up formation of irregular aggregates from submm grains.  While some of the observed trajectories during the approach of grains reflect the presence of a pure Coulomb potential, the motion is not always in agreement with pure Kepler motion. Higher-order potentials of multipole charge distributions stand as a plausible explanation for this behavior. An immediate consequence of charging is that the particles continue to stick to each other at velocities of $\sim 10 \, \rm cm/s,$ while surface forces of neutral grains are only expected to allow sticking below $\sim 1 \, \rm mm/s$. No bouncing collision was observed among hundreds of collisions in the given parameter range. Applied to early phases of planet formation, the forming aggregates are therefore the first steps in a new growth phase beyond the traditional bouncing barrier in planet formation.}
  % conclusions heading (optional), leave it empty if necessary 
   {}


   \maketitle


\section{Introduction}

With planetesimal formation in mind, we report on an experimental work in collisional charging. 
Charging in protoplanetary disks has been the subject of extensive study. The energetic radiation of various sources, whether stellar, cosmic, or due to radioactive decay, charge gas molecules regularly \citep{Cleeves2013, Bergin2007, Johansen2018}. While the ionization fraction is usually low in the optically thick inner part of the disk, this can be sufficient to drive magneto-rotational instabilities, which are often thought to be responsible for generating turbulence \citep{Balbus1991, Flock2012}. Grains placed in these environments, especially close to the surface of the disk, are charged as well in the process of their interacting with the gas ions and electrons \citep{Ilgner2012, Matthews2013, Ivlev2016}. If grain charging is biased to only one polarity, grains will experience a Coulomb repulsion that can lead to a charge barrier as grains form larger aggregates \citep{Okuzumi2009, Okuzumi2011a, Okuzumi2011b}. Closer to the surface of the disk, grains might also charge due to a direct photoelectric effect that might offset a negative bias within a plasma environment and lead to zero net charge \citep{Akimkin2015}. Locally, this allows for further aggregation without Coulomb repulsion.

This is all of little importance in the midplane of the disk, where timescales of charging and recharging by radiation might be low (i.e., the dead zone) especially compared to particle collision timescales \citep{Ueda2019, Steinpilz2020a, Jungmann2021}. We note that we are referring to collisions between solid particles here, not collisions with molecular ions or electrons. Grain-grain collisions have been discussed in the past as another way of charging grains triboelectrically, such as in the context of chondrule formation. As in the case of a thunderstorm, grains of different sizes that collide with one another and bounce off are charged differently \citep{Pahtz2010} and, typically, the small grains are charged negatively \citep{Lacks2011, Waitukaitis2014}. If size segregation occurs by size-dependent drift velocities or sedimentation speeds, small and large grains might separate, which leads to a large-scale charge separation. This can trigger lightning, which might, in turn, eventually lead to chondrule formation by melting \citep{Desch2000, Muranushi2010, Muranushi2015}. Collisional charging might go other ways, especially if colliding grains are of similar size and similar material. In that case, charging might induce a new growth phase of larger particle aggregates, which is the motivation for this work.
\citet{Steinpilz2020a} showed that collisional charging  has the potential to overcome some barriers which, so far, exhibit certain flaws in models of planet formation.

These barriers arise as the primary mechanism for particle evolution, namely, the growth in hit-and-stick collisions, breaks down. Beyond the mm-size, depending somewhat on the location within the protoplanetary disk, collisions only result in bouncing and fragmentation \citep{Blum2008, Guttler2010, Zsom2010}. It is not the case, however, that further growth in the fragmentation regime would not be possible. As small particles are disrupted in a collision, they leave material on a larger body in a sort of a snowball-like fashion \citep{Wurm2005, Meisner2013, Deckers2016}. However, this process might be too slow to grow the first set of planetesimals efficiently \citep{Windmark2012}. Also, while fragmentation is often considered as the final growth barrier in simulations, bouncing is prevalent for smaller aggregates and the fragmentation regime might not be easily achieved \citep{Zsom2010, Kelling2014, Kruss2016, Kruss2017, Demirci2017}. The size of aggregates at the fragmentation and bouncing barrier are not very different as the collision energy scales with mass. In any case, different paths have been worked out with regard to proceeding with planet formation without collisions. 

Particle traps in pressure bumps, streaming instabilities, or baroclinic instabilities are examples of mechanisms where an interaction of the solid particles with the embedding gas leads to a concentration of particles \citep{Youdin2005, Johansen2007, Squire2018, Bertrang2020, Andrews2020, Carrera2020}. If all works well, the resulting particle clouds are dense enough for gravitational collapse. This is a promising way, but it requires a certain initial particle size. This is often expressed in terms of Stokes numbers St (ratio between gas-grain coupling time and orbital time) as this might serve as  a significant hydrodynamic quantity. Simulations with St = 1 regularly lead to concentrations of solids. Under certain conditions, concentrations might also occur at Stokes numbers as low as St = 0.01 \citep{Yang2017}. 

However, from a "mechanical" point of view, absolute size matters quite a bit as millimeter aggregates (mm aggregates) collide differently from meter-bodies even if they share the same Stokes number. While this somewhat depends on the disk properties, as does the bouncing barrier, the minimum size required for hydrodynamic effects to lead the way is on the cm-scale or larger \citep{Johansen2014}. To put it differently, there are one or two orders of magnitude in size that are not accounted for between conventional collisional growth and hydrodynamic interplay. 

This is where collisional charging becomes important. Any contact between two dielectric grains comes along with charge transfer that is due to triboelectric effects. This is especially the case at the bouncing barrier, where collisions charge the grains, which can initiate a new growth phase as charges can eventually bind the colliding grains together. In a simplified picture, we may use the analogy of ionic crystals. While salt is  electrically neutral overall, the ions, with their alternating positive and negative charges, bind the crystal strongly. Indeed, \citet{lee2015} found small aggregates arranged in crystal-like structures in short duration microgravity experiments within a free-falling granular medium.

Recently, \citet{Steinpilz2020a} carried out drop tower experiments to find that submillimeter (submm) grains charged by collisions subsequently formed aggregates that consisted of hundreds of grains. Being several cm in size with constituents of submm size, they clearly demonstrate the potential of collisional charging to close the size gap in pre-planetary evolution. In these experiments, aggregates with an intermediate size of about ten grains were also observed, with a note, however, that grains within the aggregates were arranged with their net charges in a non-crystal-like structure. With an even more counterintuitive finding, \citet{Steinpilz2020a} also observed aggregates that were composed solely of grains of the same polarity. This is in agreement with simulations by \citet{Matias2018}, who found that induced dipoles can provide strong binding.  

Obviously, the net charge of individual grains is not the only parameter that is important in aggregate stability.
Beyond induced dipoles, it is also the detailed permanent charge distribution on an individual grain that might be important. Along the lines of this argument, \citet{Jungmann2018} studied the threshold velocity at which spherical grains stick to a metal plate. It strongly depended on the grain charge, but the results could only be explained if the net charge was assumed to be offset from the center of the grains. This implies that the charge is not homogeneously distributed over the surface.  The authors argue that charge and mirror charge are closer together than the centers of mass. This boosts the particle speed right before a collision and leads to a larger absolute energy loss. This keeps the particle trapped in the Coulomb potential afterwards.

In a more detailed study, \citet{Steinpilz2020b} measured large dipole moments on pairs of grains. These dipole moments could not be explained by homogeneous charge distributions. The measurements could be modeled with patches of charge on the surface of the particles. Irregularly charged surfaces are also consistent with the work by \citet{Grosjean2020} and \citet{Apodaca2010} on identical material charging.
Strong interactions between charge patches should also influence a collision between two dielectric grains and their final approach to one another if these particles have a non-homogeneous charge distribution. 

To evaluate the importance of these parts in the stability of particle aggregates. we carried out dedicated experiments in the drop tower in Bremen to follow the motion of charged grains in greater detail.
At larger distances between grains, the motion should be determined by Coulomb attraction of the two net charges and, subsequently, the particles should enter a Kepler-like orbit. Such orbits for similarly sized grains have already been observed by \citet{lee2015}.
As we use a longer microgravity duration here, we can trace the motion of the grains more closely for a longer time and study the bottom-up formation of clusters.

\section{Experiment\label{sec:experiment}}

\begin{figure}
 \centering 
\includegraphics[width=0.8\columnwidth]{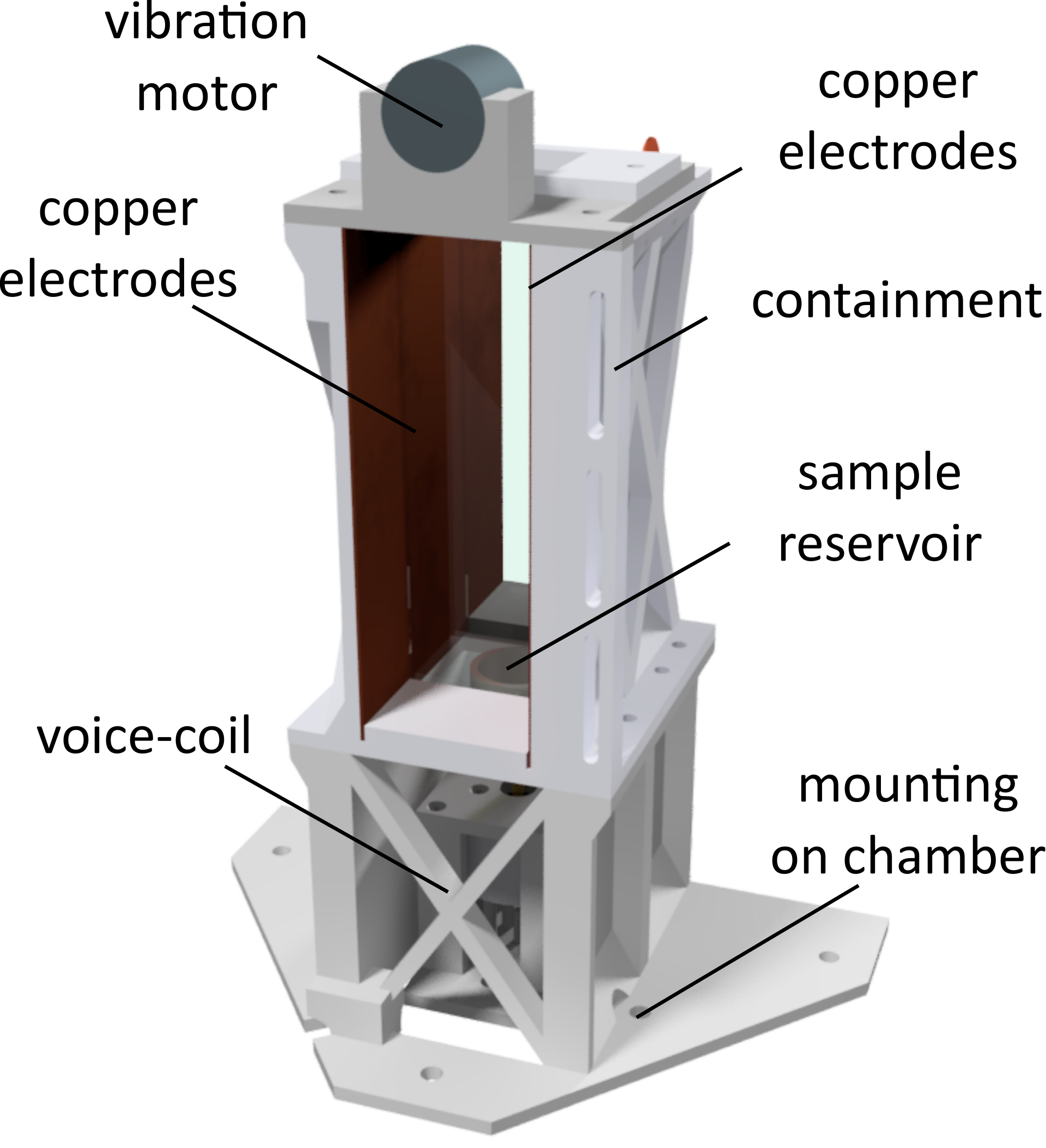}
\caption{Main components of the experiment. The illumination from the back and the camera observing from the front is not shown here. }
\label{fig:setup}
\end{figure}

The experiments were carried out in the drop tower in Bremen under microgravity. Launched from ground, the drop tower provides about 9 sec of microgravity. The setup of the experiment and its performance are similar to the experiments described in \citet{Jungmann2018} and \citet{Steinpilz2020a}.
In Fig. \ref{fig:setup}, we show the basic components. The main part is a 5 x 4.4 x 14 cm (width x depth x height) capacitor volume which can be reduced in height down to 9.5 cm. This volume is back-illuminated for a camera placed at the front to observe particles within the total volume. The 15 x 11.5 cm sized copper electrodes are placed perpendicular to the camera's perspective. They are much larger than the observation volume in order to provide a homogeneous electric field. 
In the experiments reported here, basalt spheres with a diameter of 500-600 $\mu$m and a mass of $0.26 \pm 0.04$ mg were filled into a sample reservoir which is an aluminum cylinder, coated with a layer of basalt spheres. The cylinder is open at the top with an orifice of 25 mm. Before each launch, the grains are shaken for 15 minutes by a voice-coil underneath to charge the grains. After launch, the capsule relaxes as it reaches microgravity. Due to this relaxation, the sample is slowly ejected into the free volume between the copper electrodes. At this point, grains are moving in a gravity-free environment. 

\begin{figure}
 \centering 
\includegraphics[width=0.95\columnwidth]{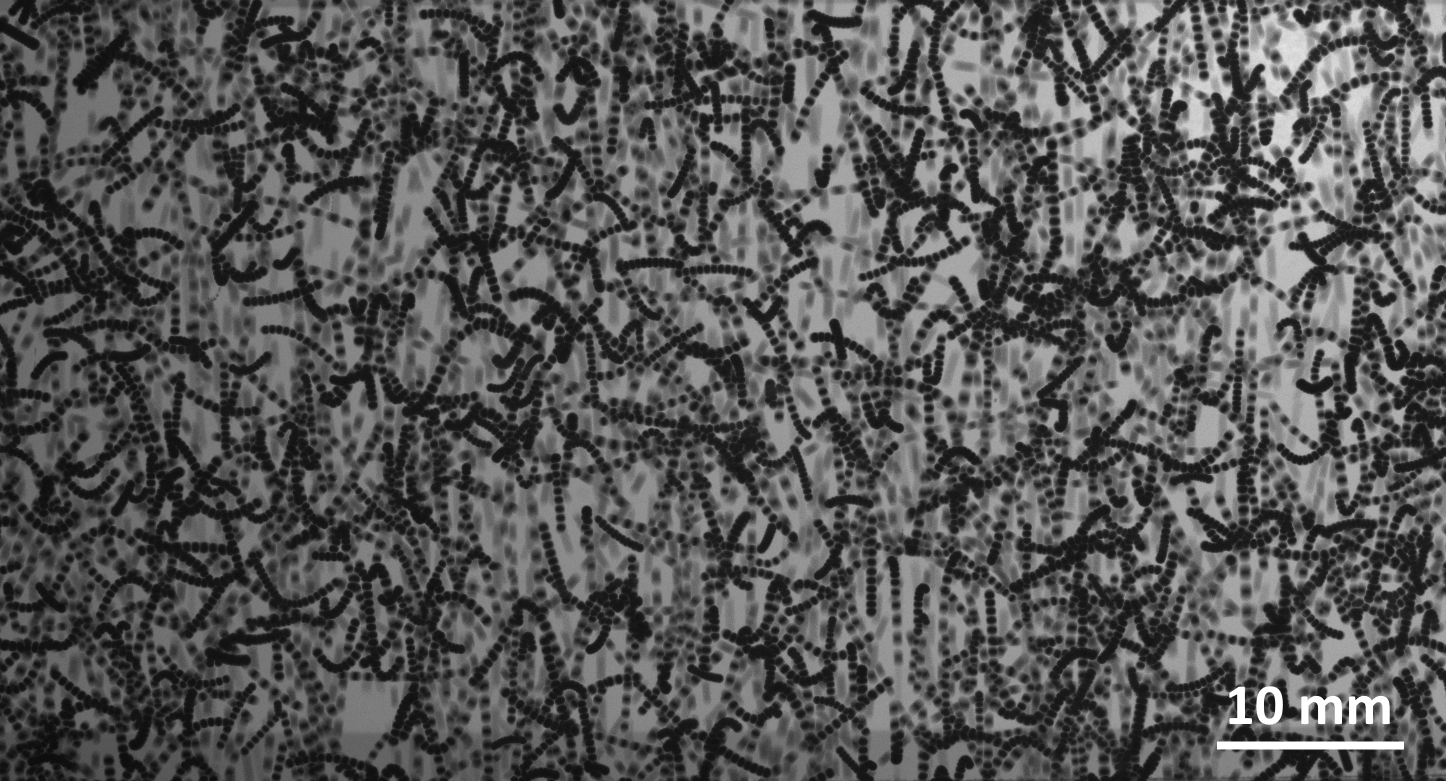}
\caption{\label{fig:waves} Overlay of ten images at the initial phase of the experiment is shown, covering a total period of 100 ms. The applied AC voltage on the capacitor (top and bottom) leads to numerous fast collisions and an oscillation of charged grains. This can be seen as curved trajectories.}
\end{figure}

In the electrical field applied to the capacitor, the grains are accelerated to the electrodes. This acceleration can be measured from the trajectories of the grains. As the mass of the grains is known, the accelerations are direct measures of the net charge on a grain. However, a constant electric field would separate positive and negative grains. This would have a strong influence on further interaction as only the grains of same polarity could continue to collide besides concentrating grains at the electrodes. To avoid these effects and, instead, generate a homogeneous mixture of grains, an AC electric field of 40 kV/m and 3 Hz is applied. This still allows us to determine net charges from the resulting oscillation but it does not manage to separate charges of different signs. Since the cloud of spheres is dense, this leads to many fast collisions distributing the grains homogeneously over the volume after about 2 sec. This phase, along with the oscillations, can be seen in an overlay of particle images in Fig. \ref{fig:waves}. At 2.5 \textbf{sec} after the onset of microgravity, the field is switched off so that the particles can move freely. Their motions are further observed by the camera with 110 frames per second. During flight, a vibration motor can slightly move the copper electrodes to release spheres sticking to the wall, but this is of minor importance here.

\begin{figure}
    \centering
{{\includegraphics[width=0.45 \columnwidth]{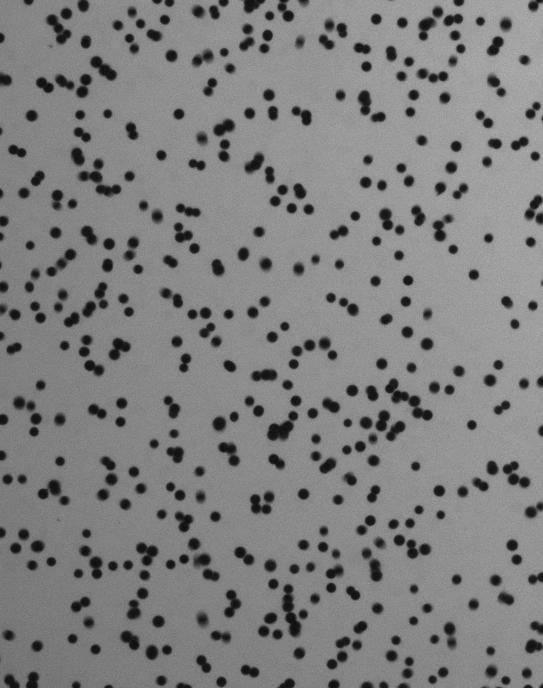} }}%
    \qquad
{{\includegraphics[width=0.45 \columnwidth]{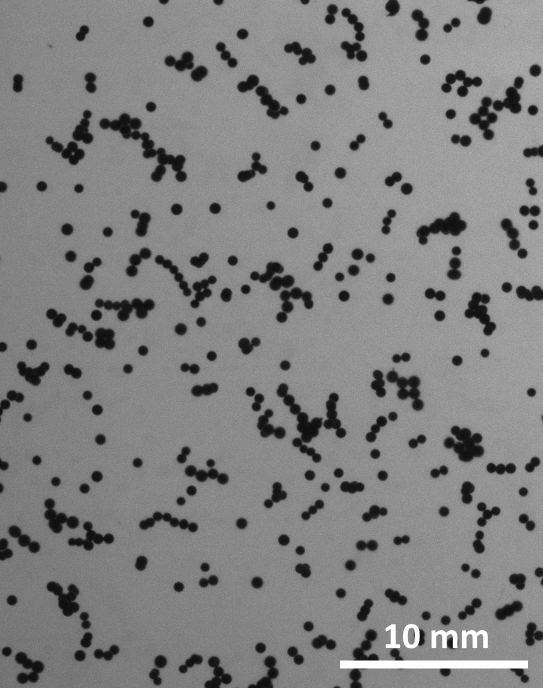} }}%
    \caption{\label{fig:cool_down}Snapshots of particle evolution. Only part of the volume is shown to enhance visual resolution. Left: 3 sec in microgravity shortly after the cloud starts to cool. Right: Cloud 9 \textbf{sec} in microgravity shortly before the experiment has fallen back on the ground and microgravity ends. A transition from individual grains to cluster formation is visible. We note that the depth of the volume is 44 mm, thus, it may be the case that particles that appear to be next to each other are actually quite distant in reality.}
\end{figure}

Once the field is switched off, the granular gas is no longer driven and it slows down due to gas drag and further collisions on timescales of several seconds. At this stage, the collisions lead to sticking. This leads to the bottom-up formation of larger clusters up to ten spheres per cluster as shown in Fig. \ref{fig:cool_down}. 

\begin{figure}
 \centering 
\includegraphics[width=0.9\columnwidth]{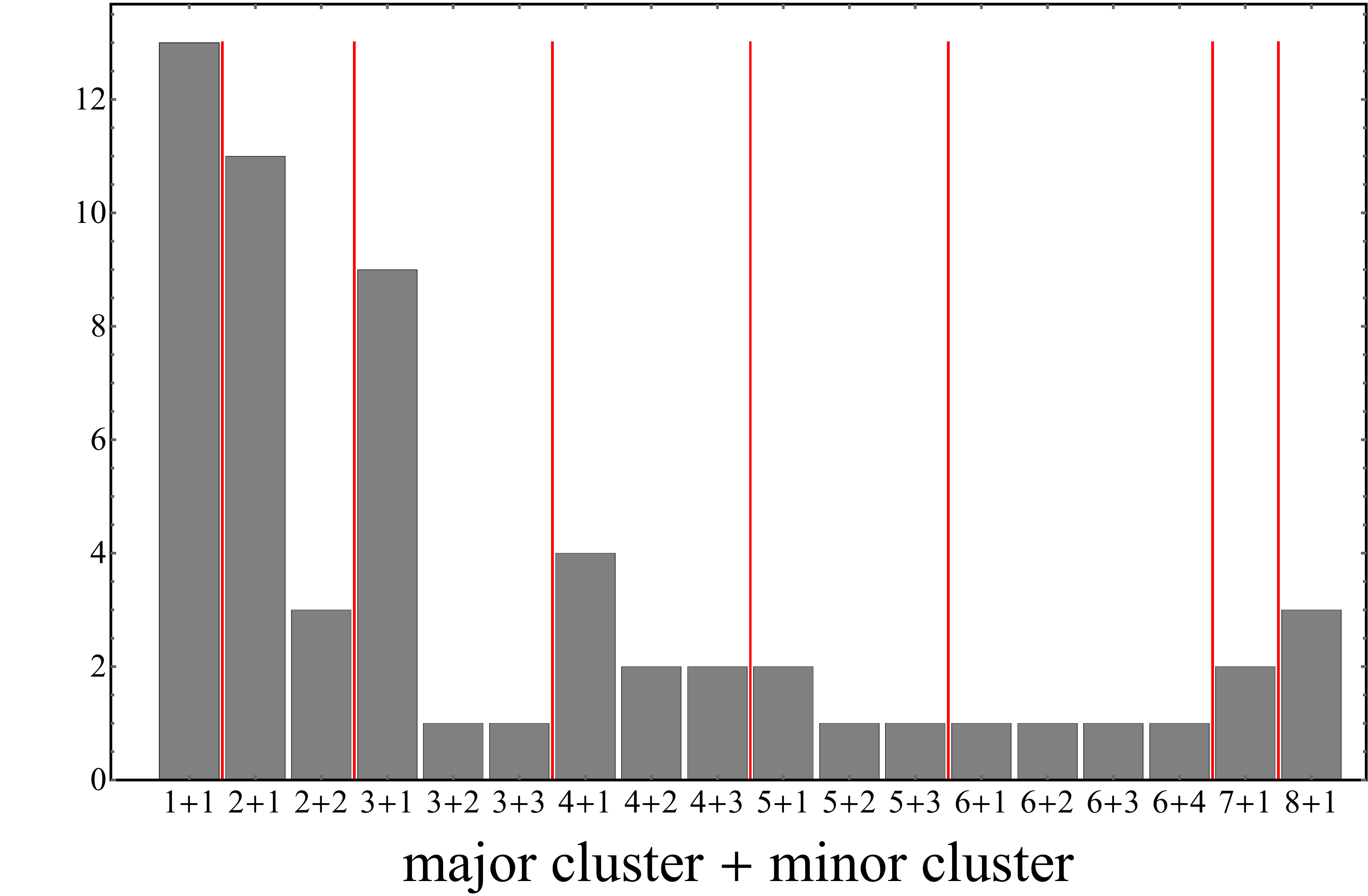}
\caption{\label{fig:colltype}Number of collisions of clusters with given grain numbers explicitly tracked. Labels are ordered by the largest cluster and denote the number of grains in the smaller cluster. The red lines mark the transitions to a larger major cluster.}
\end{figure}

\begin{figure}
 \centering 
\includegraphics[width=0.9\columnwidth]{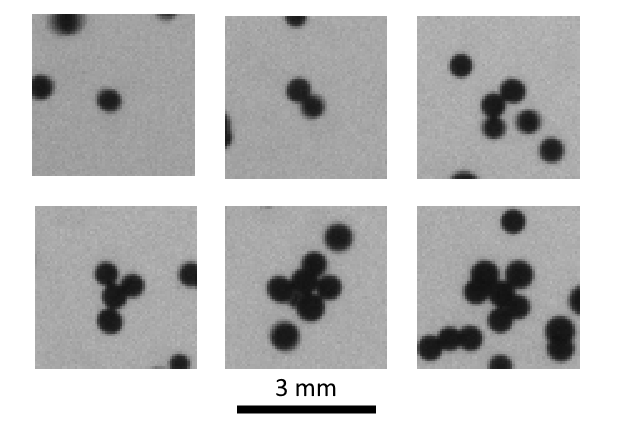}
\caption{\label{fig:6er}Formation sequence of a six-particle cluster starting with a single sphere (top left) growing by the addition of individual spheres. }
\end{figure}

\begin{figure}
 \centering 
\includegraphics[width=0.9\columnwidth]{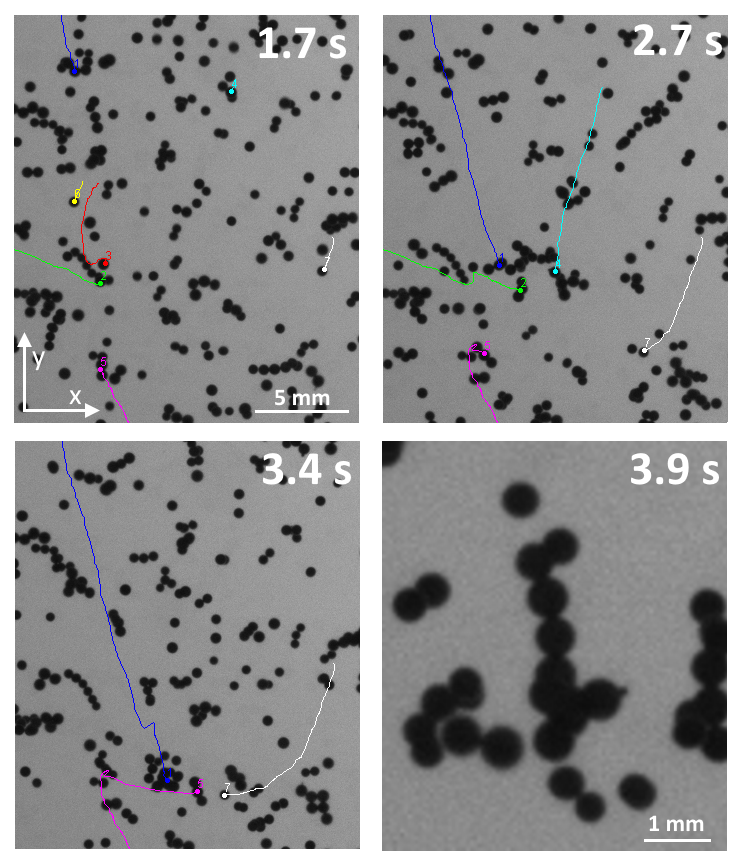}
\caption{\label{fig:10er} Largest cluster observed consists of ten particles and is shown in the bottom-right image. The other three images of this figure show part of the formation history of this cluster. The coloured lines are trajectories of individual grains or smaller clusters that collide, stick together, and form larger clusters. These clusters collide further and become the large cluster, eventually.  The time in the top right corner of each image shows the elapsed time after the beginning of the first trajectory. 
The final large cluster is formed of two trimers (cluster consisting of three individual grains) and one quad (cluster with four grains). After a collision, only one of the two trajectories is continued.
For the first trimer, sequential growth is observed (1+1+1). Afterwards, it collides and sticks to the other trimer. This second trimer grew somewhere else which is not shown in the images. The quad is observed to form from 2+1+1 particles. It then combines with the two trimers. The final cluster is shown after 3.9 sec in higher resolution.} 
\end{figure}

During the 6 sec of evolution, the formation of larger clusters can be observed explicitly. Figure\ \ref{fig:colltype} gives an overview of the combinations of collisions tracked. 
Clusters predominantly grow by adding monomers and Fig. \ref{fig:cool_down} also clearly shows that monomers are present to a large amount until the end of the experiment. Figure \ref{fig:6er} shows an example sequence centered on a growing six monomer cluster formed by adding monomers each time. Figure \ref{fig:10er} shows the formation of a ten-grain cluster (the largest one observed) including the trajectories of all participating grains. In total, the center of 13 clusters were tracked, including 59 sticking collisions. We note that all observed collisions after switching off the external electric field result in sticking. 

\begin{figure}
 \centering 
\includegraphics[width=0.9\columnwidth]{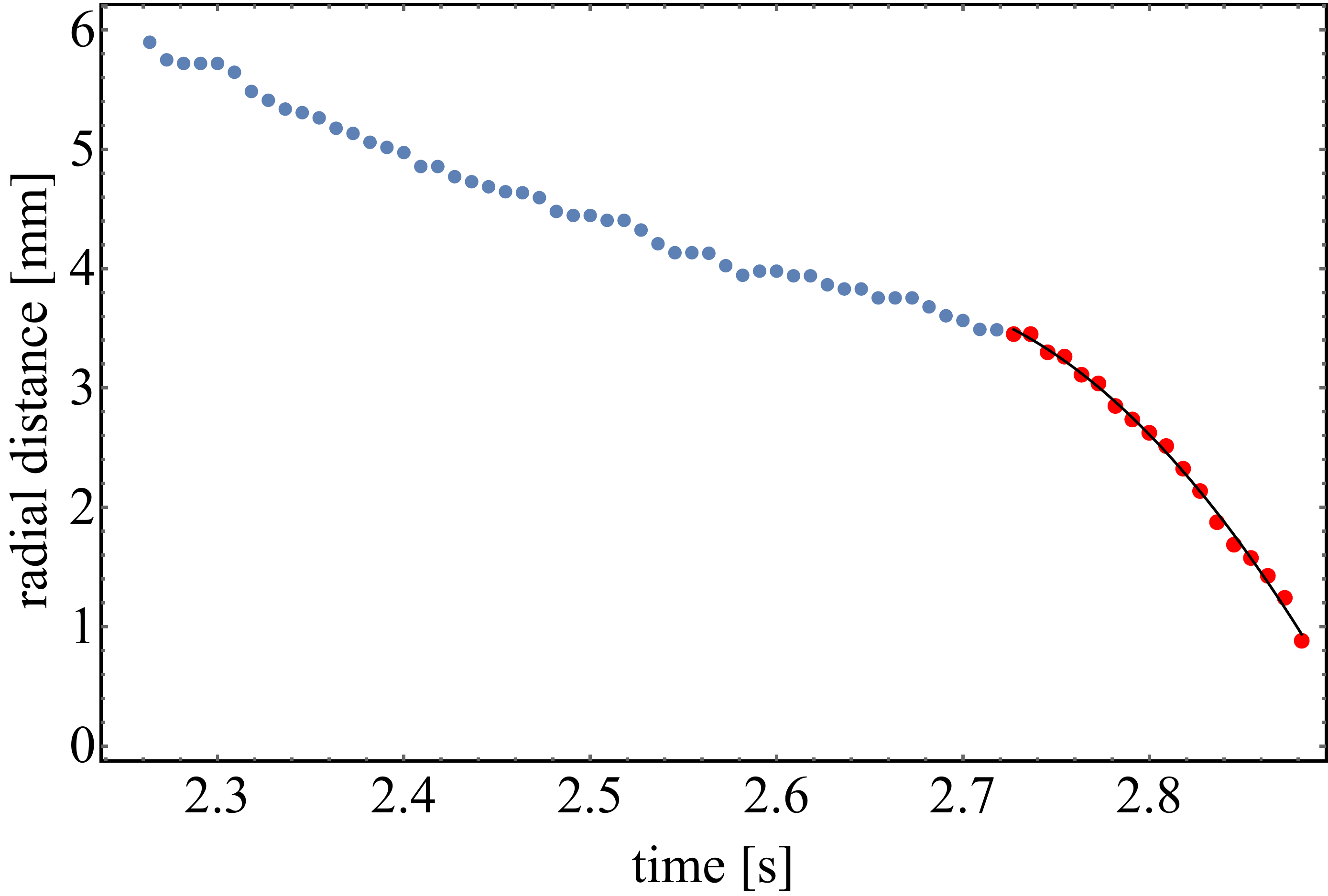}
\caption{\label{fig:parabola_fit}Example of the distance between a three-particle cluster and an approaching single sphere tracked over time. Blue dots show tracked coordinates; red dots show data used for fitting; black line: parabolic fit.}
\end{figure}

Beyond counting, certain properties of the collisions can be extracted from the tracks, that is, collision velocities and velocity profiles upon approach. Figure \ref{fig:parabola_fit} shows the approach of a grain towards a trimer over time. The motion is essentially linear in distance or force-free until 2.7 sec when it gets accelerated as a certain distance is reached. To determine the collision velocity, we used simple parabolic fits to the accelerated part of the trajectories. The region of acceleration is set  by eye.
This distance is also used as a measure of the cross-section of the electrostatic attraction, noting, however, that this is only a manual measurement within the resolution of the tracks.

\section{Results}

We can go one step further in the analysis of the tracks to evaluate the homogeneity of the surface charge distribution.

\subsection{Charges}

\begin{figure}
\centering 
\includegraphics[width=0.9\columnwidth]{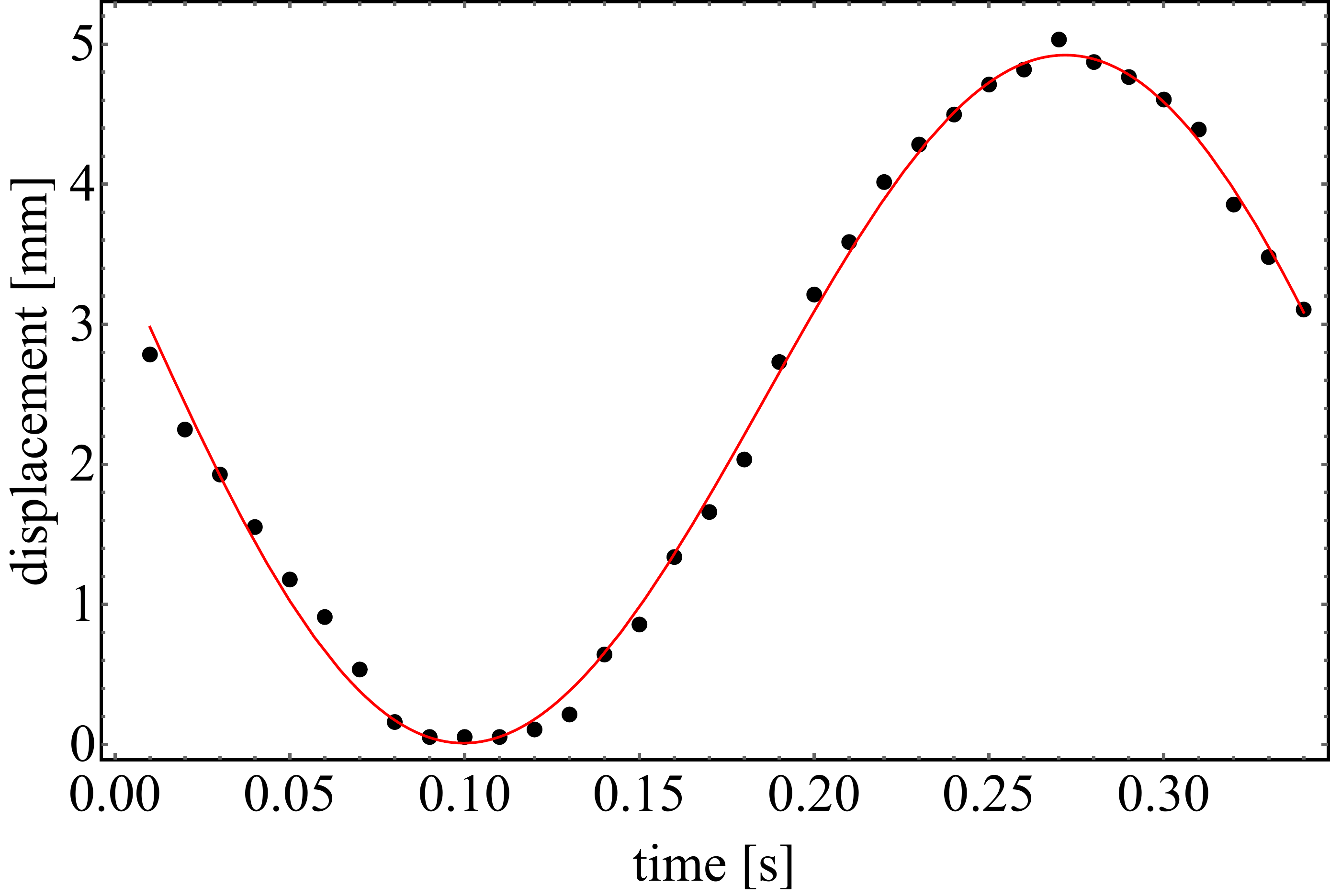}
\caption{\label{fig:wobbler-fit}
Oscillation of a single grain and fit of a sine-like movement giving a net charge of $1.7 \cdot 10^{7} e$.}
\end{figure}

Once the spheres enter the capacitor, a sine-like electric field is applied with a frequency of 3 Hz. The particles oscillate with the same frequency (Fig. \ref{fig:wobbler-fit}). Typically, the particles oscillate for one period until they collide with another particle. This is sufficient to fit an amplitude of $A_0$ and the net charge $q$ can be calculated as

\begin{equation}
    q= \dfrac{A_0\, d\, m \,\omega_0^2}{U}.
\end{equation}

Here, $U$ is the applied voltage of 4000 V, $d$ is the distance of the capacitor plates of 5 cm, $m$ is the particle mass, and $\omega_0$ is the applied (circular) frequency.

Eight particles were tracked this way to get a selection of the net charge on the grains. They range from $1 \cdot 10^7$ e to $1 \cdot 10^8$ e. We did not determine a more detailed charge distribution as this would be subject to a strong selection bias. The given range is only meant as an order of magnitude for the largest net charges in the distribution. We consider this to be sufficient in this context.

As seen in Fig. \ref{fig:parabola_fit}, there is an acceleration of the particles towards one another. Considering two monopole charges, $q_1$ and $q_2$, we would expect a velocity profile of a Coulomb potential.

\begin{equation}
    E_{\text{pot}}(r)= \dfrac{1}{4 \pi \epsilon_0} \dfrac{q_1 q_2}{r}
    \label{eq:coulomb-pot}
,\end{equation}

Using energy conservation, this can readily be calculated as

\begin{equation}
    v(r)=\sqrt{\dfrac{2}{m}\left(E - \dfrac{1}{4 \pi \epsilon_0} \dfrac{q_1 q_2}{r} \right)}
    \label{eq:v-coulomb}
,\end{equation}
where $E$ is the total energy.
To estimate the order of magnitude of the involved net charges, we can set $q=q_1=q_2$ without any loss of generality. 

\begin{figure}
 \centering 
\includegraphics[width=0.9\columnwidth]{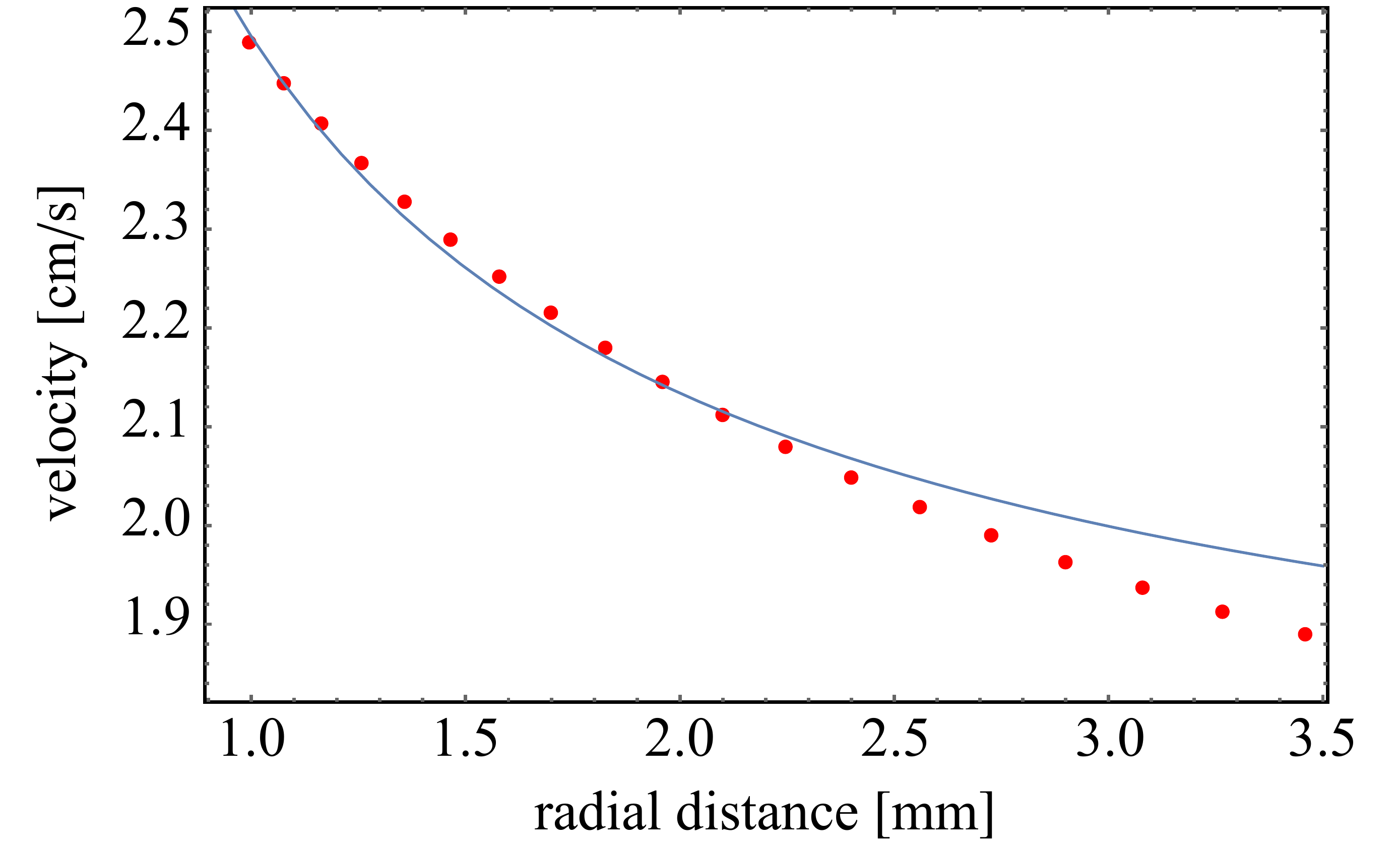}
\caption{\label{fig:v-over-r2} Velocity profile of a collision of two dimers (red) compared to a pure net charge moderated profile of two charges with plus and minus $1.4 \cdot 10^7 e $ (blue).}
\end{figure}

\begin{figure}
 \centering 
\includegraphics[width=0.9\columnwidth]{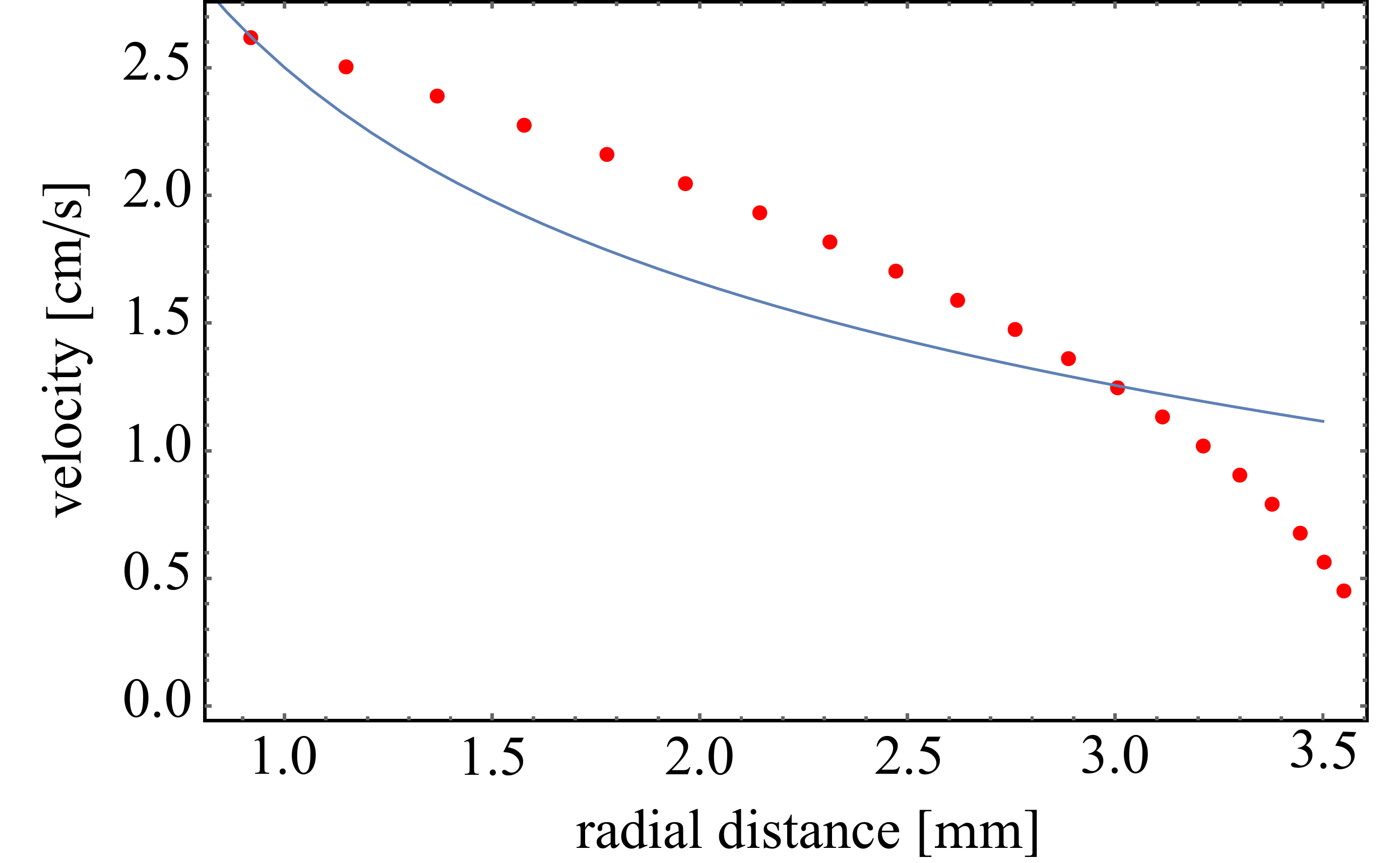}
\caption{\label{fig:v-over-r1}Velocity profile of a collision of a dimer and a single sphere (red) compared to a pure net charge moderated profile of two charges with $2 \cdot 10^7 e $. As the acceleration does not increase on approach, it cannot be simulated by net charges alone.}
\end{figure}

We show two examples of velocity profiles from Eq. \ref{eq:v-coulomb} in comparison to measured track velocities  in Figs. \ref{fig:v-over-r2} and \ref{fig:v-over-r1}. In the first case, the closest approach agrees with a simple Coulomb law. The differences at larger distances are likely due to the disturbances from the cloud. As the net charge, the fit gives $1.4 \cdot 10^7$ e, which is in agreement to the individually measured charges.

Nonetheless, many cases deviate strongly from this simplified behaviour. This is visualized in the example shown in Fig. \ref{fig:v-over-r1}. As the acceleration is not increasing upon approach, this cannot be modeled by the simple Coulomb law and, thus, other explanations are needed. The most plausible one is that the movement of the particles is not only influenced by two net charges but also by higher orders of charges, that is, dipoles and multipoles. Since clusters are composites of differently charged spheres, we expect the agglomerates to cause those higher-order fields. In addition, \citet{Steinpilz2020b} suggested having a non-homogeneous charge distribution on the particles' surface, which allows for even single spheres to have higher-order electric fields.

Unfortunately, in a rigorous sense, the data do not provide enough information to prove the assumption that higher-order electric fields are the reason for the deviations seen in the trajectories. In general, there are less than 20 available frames depicting the approach of two particles. There are other explanations at hand, such as the notion that the field of the particle cloud in the surroundings influences the trajectories. This is not likely, however, in general terms as a mean field disturbance would be low due to the overall charge neutrality of the cloud. Also, many trajectories of charged grains (see e.g., Fig. \ref{fig:parabola_fit}) are force-free unless they are close to another grain. 
The largest uncertainty is the missing third dimension, which can lead to a misinterpretation of two-dimensional trajectories when there is only a 2d energy balance that is considered. More complex orbital fitting as, for instance, in \citet{lee2015} would be required but our data do not provide the necessary accuracy.

Nevertheless, we showed at this stage that charges are heavily involved in the collision dynamics and play an important role there. 
At least, charges lead to clustering due to additional attractive forces being much stronger and longer in range than van-der-Waals forces.

\subsection{Sticking velocities}

\begin{figure}
 \centering 
\includegraphics[width=0.9\columnwidth]{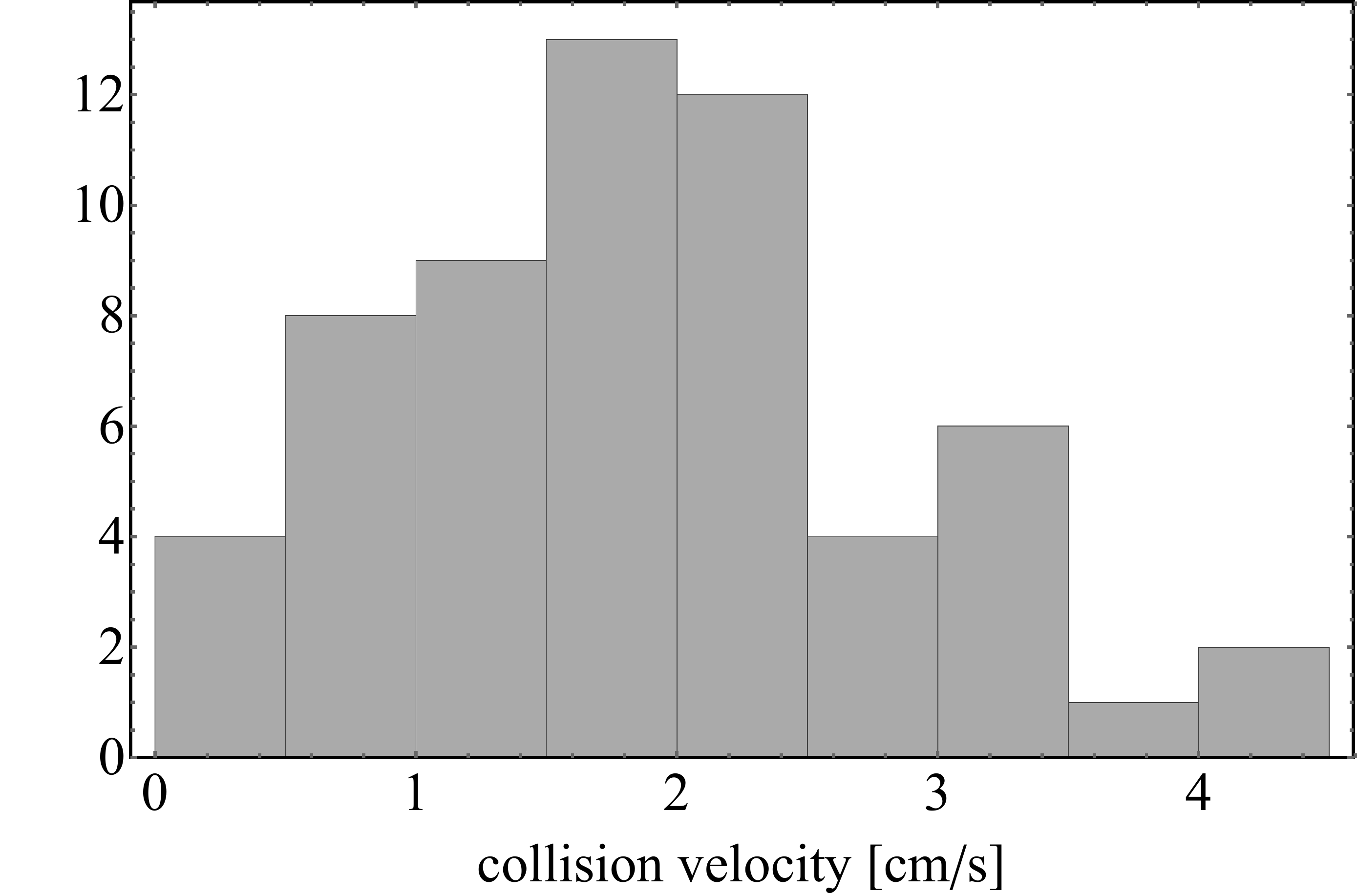}
\caption{\label{fig:k-geschw}Histogram of collision velocities measured.}
\end{figure}

Charges are expected to have an impact on the stability of aggregates and on the outcome of collisions, that is, particles should stick at higher collision velocities compared to uncharged grains.
To quantify this, in Fig. \ref{fig:k-geschw} we show a histogram of the collision velocities measured. The typical value is 2 cm/s. The fastest collisions are beyond 4 cm/s.

As mentioned earlier in this paper, all the observed collisions result in sticking. We still show the collision velocity histogram to point out that these are not special events and to demonstrate that grains regularly stick in this velocity range. Given we see no bouncing, the threshold velocity between sticking and bouncing must be higher. Besides, as we are only tracing the two-dimensional collision velocity, the true collision velocity is higher. Taking this into account, the sticking velocity is likely more on the order of 10 cm/s or beyond. 

This value is large compared to sticking velocities expected for uncharged grain. Using \citet{Thornton1998}, we get for a collision between two uncharged grains:
\begin{equation}
    v_{\text{st}}=1.84 \left( \dfrac{(\gamma/r)^5}{\rho^3 E^{\ast 2}}\right)^{\frac{1}{6}}
,\end{equation}
with $E^{\ast}=E/(1-\nu^2)$. Here, $\gamma$ is the surface energy, $E$ is Young's modulus, $\nu$ is the Poisson ratio, and $\rho$ is the density of the material. For basaltic spheres, we use $\gamma= 0.07 \,\text{J}/\text{m}^2$ \citep{Bogdan2020}, $E=4 \cdot 10^{10}\, \text{Pa}$ \citep{Deak2009}, $\nu=0.25$ \citep{Schultz1995} and $\rho=2900\, \text{kg}/\text{m}^3$ and get    $v_{\text{st}}= 1\, \text{mm}/\text{s}$.
Therefore, the maximum measured sticking velocity is about two orders of magnitude higher than this theoretical value. The charge can exceed the attraction of cohesion by far and increase the maximum sticking velocity significantly.

It must be noted that we used solid particles to simulate mm-dust aggregates. The sticking of dust aggregates is different, however, as contacts with shifting micrometer grains are more variable and there might be no sharp threshold between bouncing and sticking. \citet{Weidling2012} studied collisions of uncharged mm-sized dust aggregates in microgravity. While they found a small fraction (7/125 collisions) sticking at velocities up to 3 cm/s (comparable to values observed in this work), most rebounded and they modeled 100 \% sticking to occur at velocities on the order of 0.1 mm/s. Here, we observed 100 \% sticking at up to 4 cm/s and beyond. 
To be more specific, \citet{Kothe2013} modeled the transition velocity dependence on particle mass by
$v_{st} \sim (m/4.5)^{0.75}$. For a dust aggregate corresponding to our basalt grains, this would be in agreement with their earlier results in \citet{Weidling2012}. 
Constituent grains in a compact aggregate have little capabilities to move and restructure the aggregate. Therefore, energy dissipation is reduced once the aggregate reaches this compact state and collisions become more elastic and it is harder to form clusters of aggregates. However, once aggregates stick to one another and form larger clusters, these clusters might restructure in collisions between clusters and clusters. This introduces a new channel for dissipating collisional energy. In agreement with this, \citet{Kothe2013} also observed sticking in cluster-cluster collisions, which seems to be possible at higher collision speeds but due to the probabilistic nature of the transition between sticking and bouncing, more data would be needed here to compare this quantitatively to single aggregate collisions.

As far as a 100\% sticking threshold is concerned, single dust aggregates do not behave better compared to compact grains, that is, they are not stickier than the compact grains we used. Due to the variability, there are sticking events comparable in velocity to the charged cases studied here. This does not mean that a few dust aggregates sticking at high velocities solve the bouncing barrier problem or are equivalent to charged grains sticking because there is a fundamental difference, which we explain below. 

Neutral dust aggregates only stick together due to short range van-der-Waals interaction or similar surface force. The contacts formed in sticking collisions of dust aggregates are not very stable. Several works by, for instance, \citet{Kelling2014}, \citet{Kruss2016}, or \citet{Kruss2017} showed that these kinds of rated breakpoints do not allow continuous growth if collision velocities remain high in contrast, for example, to a cooling granular gas. Clusters of aggregates are destroyed again eventually, as, for example, shown by the same studies \citep{Kelling2014, Kruss2017, Demirci2017}. In contrast, charge moderated sticking is long-ranged as can be seen by the increased cross-section (see Section \ref{ch:cross}) and often clusters are observed to rearrange without getting destroyed or are observed to bounce off but return to each other. This is readily observed for the 13 clusters analyzed in this paper. This is not explicitly visualized here but examples of how  charged clusters are rearranged upon collision without destruction can be found, for instance, in \citet{Steinpilz2020a}.

\subsection{\label{ch:cross} Collisional cross-sections}

For uncharged grains, the collisional cross section of two particles is that of two hard spheres which is $\pi d^2/4$ for equal sized grains of diameter $d$.
With long-ranged Coulomb interaction, the collisional cross section is altered significantly. Especially, grains charged with a different polarity attract each other and collide more readily with strongly enhanced cross sections. As we cannot measure cross-sections directly due to the 2d limitations, we quantify an increased cross-section due to charge by giving a (2d) interaction length. This is defined as the distance once attraction becomes visible in the trajectories (Fig. \ref{fig:parabola_fit}).

\begin{figure}
 \centering 
\includegraphics[width=0.9\columnwidth]{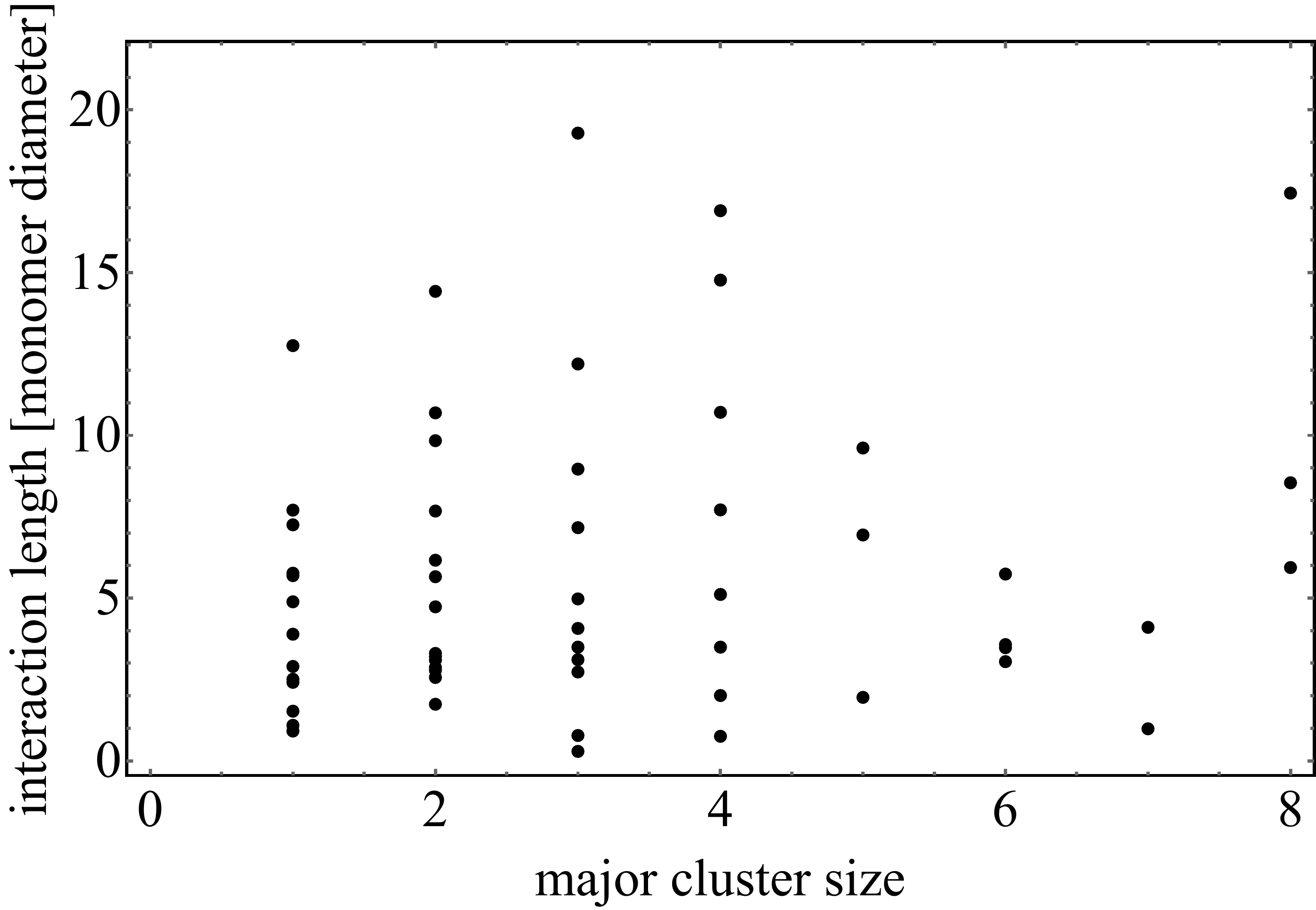}
\caption{\label{fig:interaction-length2} Interaction length over the number of grains in the major cluster of a collision.}
\end{figure}

Figure \ref{fig:interaction-length2} shows the interaction lengths determined.
The average interaction length (mean of 59 interaction lengths) is six particle diameters in our case.
One might expect that the interaction length decreases with the cluster size as positive and negative charges in a cluster neutralize. However, as seen in Fig. \ref{fig:interaction-length2} (at least up to the largest clusters hitherto observed), the interaction length can remain high. If there is a systematic trend, it is not statistically significant in our sample. Also, a further division of the data in small and large cluster influence is beyond this work. However, previous experiments by \citet{Steinpilz2020a} or \citet{Teiser2021} also show that aggregates consisting of many particles are not  neutral overall but they still carry a certain net charge. Aggregates do not become neutral.

\section{Conclusion}

For the first time, we directly observed a cloud of submm-sized, collisionally charged particles evolving into aggregates. Stable aggregates are formed by sticking collisions in a period of a few seconds. The basaltic particles move freely under microgravity and attract one another due to their charged nature. 
The long-range influence of the net charges is noticeable in an enhanced collisional cross-section. Sticking velocities are higher compared to neutral grains.  Charge is clearly a boost factor in aggregation. 

This observation complements the work of \citet{Jungmann2018}, who have already showed that the sticking velocity threshold increases strongly in collisions of a charged particle with a copper wall. Here, we demonstrate that this also applies to particle-particle collisions. Higher multipole moments of a complex charge pattern \citep{Steinpilz2020b} may also play an important role. For two particles in contact, the local charge at the contact point matters the most, due to small distances, which can act as a powerful glue for a cluster of particles. Particle trajectories recorded in our experiments point in this direction but due to limitations of the experiment, that is, 2d trajectories, the data are non-conclusive here. Our results hint at the fact that non-homogeneous charge distributions in large aggregates are important, which should be further investigated in future experiments. In any case, charges are responsible for the $100\%$ sticking rate of half a mm-sized particles and aggregates thereof in collisions in the cm/s range.

In an uncharged part of protoplanetary disks, the bouncing barrier is a severe obstacle to planet formation \citep{Zsom2010}. However, bouncing charges grains even in slightly discharging protoplanetary environments \citep{Jungmann2021}. This shifts the threshold between sticking and bouncing. A charge-moderated growth phase is possible then, which could grow particles right into the regime of streaming instabilities or other particle concentrating mechanisms. \citet{Teiser2021} showed that the formation of giant clusters is possible in high-density charged  particle clouds under agitation. We complement that work in this paper by showing that individual submm grains can start a charge-induced growth up from the bottom of individual grains.\\

\section*{Acknowledgements}
This project is supported by DLR Space Administration with funds provided by the Federal Ministry for Economic Affairs and Energy (BMWi) under grant numbers DLR 50 WM 1762 and 50 WM 2142. Our thanks go to Tabea Bogdan and Tobias Steinpilz for their support at the drop tower. We appreciate the constructive comments of the two anonymous reviewers, which improved the paper strongly.\\

\bibliographystyle{aa}
\bibliography{bib}

\begin{thebibliography}{53}
\expandafter\ifx\csname natexlab\endcsname\relax\def\natexlab#1{#1}\fi

\bibitem[{{Akimkin}(2015)}]{Akimkin2015}
{Akimkin}, V.~V. 2015, Astronomy Reports, 59, 747

\bibitem[{{Andrews}(2020)}]{Andrews2020}
{Andrews}, S.~M. 2020, arXiv e-prints, arXiv:2001.05007

\bibitem[{Apodaca {et~al.}(2010)Apodaca, Wesson, Bishop, Ratner, \&
  Grzybowski}]{Apodaca2010}
Apodaca, M., Wesson, P., Bishop, K., Ratner, M., \& Grzybowski, B. 2010,
  Angewandte Chemie International Edition, 49, 946

\bibitem[{{Balbus} \& {Hawley}(1991)}]{Balbus1991}
{Balbus}, S.~A. \& {Hawley}, J.~F. 1991, \apj, 376, 214

\bibitem[{{Bergin} {et~al.}(2007){Bergin}, {Aikawa}, {Blake}, \& {van
  Dishoeck}}]{Bergin2007}
{Bergin}, E.~A., {Aikawa}, Y., {Blake}, G.~A., \& {van Dishoeck}, E.~F. 2007,
  in Protostars and Planets V, ed. B.~{Reipurth}, D.~{Jewitt}, \& K.~{Keil},
  751

\bibitem[{{Bertrang} {et~al.}(2020){Bertrang}, {Flock}, {Keppler}, {Trifonov},
  {Penzlin}, {Avenhaus}, {Henning}, \& {Montesinos}}]{Bertrang2020}
{Bertrang}, G. H.~M., {Flock}, M., {Keppler}, M., {et~al.} 2020, arXiv
  e-prints, arXiv:2007.11565

\bibitem[{Blum \& Wurm(2008)}]{Blum2008}
Blum, J. \& Wurm, G. 2008, Annu. Rev. Astron. Astrophys., 46, 21

\bibitem[{{Bogdan, T.} {et~al.}(2020){Bogdan, T.}, {Pillich, C.}, {Landers,
  J.}, {Wende, H.}, \& {Wurm, G.}}]{Bogdan2020}
{Bogdan, T.}, {Pillich, C.}, {Landers, J.}, {Wende, H.}, \& {Wurm, G.} 2020,
  A\&A, 638, A151

\bibitem[{{Carrera} {et~al.}(2020){Carrera}, {Simon}, {Li}, {Kretke}, \&
  {Klahr}}]{Carrera2020}
{Carrera}, D., {Simon}, J.~B., {Li}, R., {Kretke}, K.~A., \& {Klahr}, H. 2020,
  arXiv e-prints, arXiv:2008.01727

\bibitem[{{Cleeves} {et~al.}(2013){Cleeves}, {Adams}, \&
  {Bergin}}]{Cleeves2013}
{Cleeves}, L.~I., {Adams}, F.~C., \& {Bergin}, E.~A. 2013, \apj, 772, 5

\bibitem[{De{\'a}k \& Czig{\'a}ny(2009)}]{Deak2009}
De{\'a}k, T. \& Czig{\'a}ny, T. 2009, Textile Research Journal, 79, 645

\bibitem[{Deckers \& Teiser(2016)}]{Deckers2016}
Deckers, J. \& Teiser, J. 2016, Monthly Notices of the Royal Astronomical
  Society, 456, 4328

\bibitem[{Demirci {et~al.}(2017)Demirci, Teiser, Steinpilz, Landers, Salamon,
  Wende, \& Wurm}]{Demirci2017}
Demirci, T., Teiser, J., Steinpilz, T., {et~al.} 2017, The Astrophysical
  Journal, 846, 48

\bibitem[{{Desch} \& {Cuzzi}(2000)}]{Desch2000}
{Desch}, S.~J. \& {Cuzzi}, J.~N. 2000, \icarus, 143, 87

\bibitem[{{Flock} {et~al.}(2012){Flock}, {Henning}, \& {Klahr}}]{Flock2012}
{Flock}, M., {Henning}, T., \& {Klahr}, H. 2012, \apj, 761, 95

\bibitem[{{Grosjean} {et~al.}(2020){Grosjean}, {Wald}, {Sobarzo}, \&
  {Waitukaitis}}]{Grosjean2020}
{Grosjean}, G., {Wald}, S., {Sobarzo}, J.~C., \& {Waitukaitis}, S. 2020,
  Physical Review Materials, 4, 082602

\bibitem[{G{\"u}ttler {et~al.}(2010)G{\"u}ttler, Blum, Zsom, Ormel, \&
  Dullemond}]{Guttler2010}
G{\"u}ttler, C., Blum, J., Zsom, A., Ormel, C.~W., \& Dullemond, C.~P. 2010,
  Astronomy \& Astrophysics, 513, A56

\bibitem[{{Ilgner}(2012)}]{Ilgner2012}
{Ilgner}, M. 2012, \aap, 538, A124

\bibitem[{{Ivlev} {et~al.}(2016){Ivlev}, {Akimkin}, \& {Caselli}}]{Ivlev2016}
{Ivlev}, A.~V., {Akimkin}, V.~V., \& {Caselli}, P. 2016, \apj, 833, 92

\bibitem[{{Johansen} {et~al.}(2014){Johansen}, {Blum}, {Tanaka}, {Ormel},
  {Bizzarro}, \& {Rickman}}]{Johansen2014}
{Johansen}, A., {Blum}, J., {Tanaka}, H., {et~al.} 2014, Protostars and Planets
  VI, 547

\bibitem[{Johansen {et~al.}(2007)Johansen, Oishi, Mac~Low, Klahr, Henning, \&
  Youdin}]{Johansen2007}
Johansen, A., Oishi, J.~S., Mac~Low, M.-M., {et~al.} 2007, Nature, 448, 1022

\bibitem[{{Johansen} \& {Okuzumi}(2018)}]{Johansen2018}
{Johansen}, A. \& {Okuzumi}, S. 2018, \aap, 609, A31

\bibitem[{Jungmann {et~al.}(2021)Jungmann, Bila, Kleinert, M{\"o}lleken,
  M{\"o}ller, Schmidt, Schneider, Teiser, Utzat, Volkenborn,
  {et~al.}}]{Jungmann2021}
Jungmann, F., Bila, T., Kleinert, L., {et~al.} 2021, Icarus, 355, 114127

\bibitem[{{Jungmann} {et~al.}(2018){Jungmann}, {Steinpilz}, {Teiser}, \&
  {Wurm}}]{Jungmann2018}
{Jungmann}, F., {Steinpilz}, T., {Teiser}, J., \& {Wurm}, G. 2018, Journal of
  Physics Communications, 2, 095009

\bibitem[{Kelling {et~al.}(2014)Kelling, Wurm, \& K{\"o}ster}]{Kelling2014}
Kelling, T., Wurm, G., \& K{\"o}ster, M. 2014, The Astrophysical Journal, 783,
  111

\bibitem[{{Kothe} {et~al.}(2013){Kothe}, {Blum}, {Weidling}, \&
  {G{\"u}ttler}}]{Kothe2013}
{Kothe}, S., {Blum}, J., {Weidling}, R., \& {G{\"u}ttler}, C. 2013, \icarus,
  225, 75

\bibitem[{{Kruss} {et~al.}(2016){Kruss}, {Demirci}, {Koester}, {Kelling}, \&
  {Wurm}}]{Kruss2016}
{Kruss}, M., {Demirci}, T., {Koester}, M., {Kelling}, T., \& {Wurm}, G. 2016,
  \apj, 827, 110

\bibitem[{Kruss {et~al.}(2017)Kruss, Teiser, \& Wurm}]{Kruss2017}
Kruss, M., Teiser, J., \& Wurm, G. 2017, Astronomy \& Astrophysics, 600, A103

\bibitem[{{Lacks} \& {Mohan Sankaran}(2011)}]{Lacks2011}
{Lacks}, D.~J. \& {Mohan Sankaran}, R. 2011, Journal of Physics D Applied
  Physics, 44, 453001

\bibitem[{Lee {et~al.}(2015)Lee, Waitukaitis, Miskin, \& Jaeger}]{lee2015}
Lee, V., Waitukaitis, S.~R., Miskin, M.~Z., \& Jaeger, H.~M. 2015, Nature
  Physics, 11, 733

\bibitem[{Matias {et~al.}(2018)Matias, Shinbrot, \& Ara{\'u}jo}]{Matias2018}
Matias, A., Shinbrot, T., \& Ara{\'u}jo, N. 2018, Physical Review E, 98, 062903

\bibitem[{{Matthews} {et~al.}(2013){Matthews}, {Shotorban}, \&
  {Hyde}}]{Matthews2013}
{Matthews}, L.~S., {Shotorban}, B., \& {Hyde}, T.~W. 2013, \apj, 776, 103

\bibitem[{Meisner {et~al.}(2013)Meisner, Wurm, Teiser, \&
  Schywek}]{Meisner2013}
Meisner, T., Wurm, G., Teiser, J., \& Schywek, M. 2013, Astronomy \&
  Astrophysics, 559, A123

\bibitem[{{Muranushi}(2010)}]{Muranushi2010}
{Muranushi}, T. 2010, \mnras, 401, 2641

\bibitem[{{Muranushi} {et~al.}(2015){Muranushi}, {Akiyama}, {Inutsuka},
  {Nomura}, \& {Okuzumi}}]{Muranushi2015}
{Muranushi}, T., {Akiyama}, E., {Inutsuka}, S.-i., {Nomura}, H., \& {Okuzumi},
  S. 2015, \apj, 815, 84

\bibitem[{{Okuzumi}(2009)}]{Okuzumi2009}
{Okuzumi}, S. 2009, \apj, 698, 1122

\bibitem[{{Okuzumi} {et~al.}(2011{\natexlab{a}}){Okuzumi}, {Tanaka},
  {Takeuchi}, \& {Sakagami}}]{Okuzumi2011a}
{Okuzumi}, S., {Tanaka}, H., {Takeuchi}, T., \& {Sakagami}, M.-a.
  2011{\natexlab{a}}, \apj, 731, 95

\bibitem[{{Okuzumi} {et~al.}(2011{\natexlab{b}}){Okuzumi}, {Tanaka},
  {Takeuchi}, \& {Sakagami}}]{Okuzumi2011b}
{Okuzumi}, S., {Tanaka}, H., {Takeuchi}, T., \& {Sakagami}, M.-a.
  2011{\natexlab{b}}, \apj, 731, 96

\bibitem[{P{\"a}htz {et~al.}(2010)P{\"a}htz, Herrmann, \& Shinbrot}]{Pahtz2010}
P{\"a}htz, T., Herrmann, H.~J., \& Shinbrot, T. 2010, Nature Physics, 6, 364

\bibitem[{Schultz(1995)}]{Schultz1995}
Schultz, R. 1995, Rock Mechanics and Rock Engineering, 28, 1

\bibitem[{Squire \& Hopkins(2018)}]{Squire2018}
Squire, J. \& Hopkins, P.~F. 2018, Monthly Notices of the Royal Astronomical
  Society, 477, 5011

\bibitem[{Steinpilz {et~al.}(2020a)Steinpilz, Joeris, Jungmann, Wolf, Brendel,
  Teiser, Shinbrot, \& Wurm}]{Steinpilz2020a}
Steinpilz, T., Joeris, K., Jungmann, F., {et~al.} 2020a, Nature Physics, 16,
  225

\bibitem[{{Steinpilz} {et~al.}(2020b){Steinpilz}, {Jungmann}, {Teiser}, \&
  {Wurm}}]{Steinpilz2020b}
{Steinpilz}, T., {Jungmann}, F., {Teiser}, J., \& {Wurm}, G. 2020b, New Journal
  of Physics

\bibitem[{Teiser {et~al.}(2021)Teiser, Kruss, Jungmann, \& Wurm}]{Teiser2021}
Teiser, J., Kruss, M., Jungmann, F., \& Wurm, G. 2021, The Astrophysical
  Journal Letters, 908, L22

\bibitem[{Thornton \& Ning(1998)}]{Thornton1998}
Thornton, C. \& Ning, Z. 1998, Powder technology, 99, 154

\bibitem[{{Ueda} {et~al.}(2019){Ueda}, {Flock}, \& {Okuzumi}}]{Ueda2019}
{Ueda}, T., {Flock}, M., \& {Okuzumi}, S. 2019, \apj, 871, 10

\bibitem[{{Waitukaitis} {et~al.}(2014){Waitukaitis}, {Lee}, {Pierson},
  {Forman}, \& {Jaeger}}]{Waitukaitis2014}
{Waitukaitis}, S.~R., {Lee}, V., {Pierson}, J.~M., {Forman}, S.~L., \&
  {Jaeger}, H.~M. 2014, \prl, 112, 218001

\bibitem[{{Weidling} {et~al.}(2012){Weidling}, {G{\"u}ttler}, \&
  {Blum}}]{Weidling2012}
{Weidling}, R., {G{\"u}ttler}, C., \& {Blum}, J. 2012, \icarus, 218, 688

\bibitem[{Windmark {et~al.}(2012)Windmark, Birnstiel, G{\"u}ttler, Blum,
  Dullemond, \& Henning}]{Windmark2012}
Windmark, F., Birnstiel, T., G{\"u}ttler, C., {et~al.} 2012, Astronomy \&
  Astrophysics, 540, A73

\bibitem[{Wurm {et~al.}(2005)Wurm, Paraskov, \& Krauss}]{Wurm2005}
Wurm, G., Paraskov, G., \& Krauss, O. 2005, Icarus, 178, 253

\bibitem[{Yang {et~al.}(2017)Yang, Johansen, \& Carrera}]{Yang2017}
Yang, C.-C., Johansen, A., \& Carrera, D. 2017, Astronomy \& Astrophysics, 606,
  A80

\bibitem[{Youdin \& Goodman(2005)}]{Youdin2005}
Youdin, A.~N. \& Goodman, J. 2005, The Astrophysical Journal, 620, 459

\bibitem[{{Zsom} {et~al.}(2010){Zsom}, {Ormel}, {G{\"u}ttler}, {Blum}, \&
  {Dullemond}}]{Zsom2010}
{Zsom}, A., {Ormel}, C.~W., {G{\"u}ttler}, C., {Blum}, J., \& {Dullemond},
  C.~P. 2010, \aap, 513, A57

\end{thebibliography}

\end{document}